# Porous nanostructured metal oxides synthesized through atomic layer deposition on a carbonaceous template followed by calcination

Shaoren Deng,[‡, a] Mert Kurttepeli,[‡, b] Daire J. Cott,[c] Sara Bals[b] and Christophe Detavernier[a, *]

Porous metal oxides with nano-sized features attracted intensive interest in recent decades due to their high surface area which is essential for many applications, e.g. Li ion batteries, photocatalysts, fuel cells and dye-sensitized solar cells. Various approaches were so far investigated to synthesize porous nanostructured metal oxides, including self-assembly and template-assisted synthesis. For the latter approach, forests of carbon nanotubes are considered as particularly promising templates, with respect to their one dimensional nature and the resulting high surface area. In this work, we systematically investigate the formation of porous metal oxides ($Al_2O_3$, $TiO_2$, $V_2O_5$ and ZnO) with different morphologies using atomic layer deposition on multi-walled carbon nanotubes followed by post deposition calcination. X-ray diffraction, scanning electron microscopy accompanied with X-ray energy dispersive spectroscopy and transmission electron microscopy were used for the investigation of morphological and structural transitions at the micro- and nano-scale during the calcination process. The crystallization temperature and the surface coverage of the metal oxides and the oxidation temperature of the carbon nanotubes were found to produce significant influence on the final morphology.

## Introduction

Porous metal oxide nanostructures with high surface area are attractive for photocatalysis, dye-sensitized solar cells and lithium ion batteries applications and have been investigated intensively in recent years.[1-3] Atomic layer deposition (ALD) is a powerful tool to deposit ultrathin metal oxide films of high quality and offers the possibility of decorating porous materials with very thin coatings due to its self-limited reaction mechanism.[4] Various porous nanostructures can therefore be synthesized through ALD onto different porous templates such as peptide, anodized alumina, nanofibrillar aerogels, mesoporous block copolymer and multi-walled carbon nanotubes (MWCNTs). After template removal, $TiO_2$ nanotubes, $RuO_2$ nanotubes, Pt nanotubes, interconnected hollow metal oxide nanotubes and 3D network of Pt nanowires were successfully synthesized.[5-10] Usually, the removal of the CNTs results in hollow metal oxide tubes, as expected based on the uniform and conformal nature of the ALD technique. However, in our previous work, we demonstrated the synthesis of photocatalytically active $TiO_2$ nanoparticle chains forests by coating MWCNTs with $TiO_2$ using ALD and subsequent calcination to remove the carbon.[11] The resulting material offered a high photocatalytical activity towards the degradation of acetaldehyde under UV light.[12] Such morphological transformation from a tubular coating deposited onto MWCNTs into the chains of $TiO_2$ nanoparticles is dramatically different from the expected morphology of hollow oxide nanotubes and therefore warrants further investigation. In this work, we systematically studied the calcination of ALD-coated vertical aligned MWCNTs and investigated the kinetics of the carbon removal and the morphological and structural changes at the micro- and nano-scale before and after calcination.

## Experimental section

MWCNTs were synthesized by plasma enhanced chemical vapor deposition (PECVD). 70nm TiN and 1nm Co as catalyst for MWCNT growth were sputtered on 8-inch Si wafer (Endura PVD tool, Applied Materials, USA). A 5 minutes $NH_3$ plasma treatment for the Co transformed the film into active metal nanoparticles to grow MWCNTs. ~6.5μm MWCNTs were grown in a microwave (2.45 GHz) PECVD (TEL, Japan) by using a $C_2H_4/H_2$ mixture gas at a temperature >550$^o$ C for 30mins. The MWCNTs were then loaded into a homemade ALD system with a base pressure of $2*10^{-7}$ mbar. Metalorganic precursors and ozone were alternatively pulsed into the chamber while the MWCNTs sample was heated up to 100$^o$C for $TiO_2$, ZnO and $Al_2O_3$ and 150$^o$C for $VO_x$. Tetrakis-(dimehtylamido) titanium (TDMAT) (99.9%, Sigma Aldrich), Tetrakis-(ethylmethylamino) vanadium (TEMAV) (Air Liquide), Diethylzinc (DEZn) (≥52 Wt%, Sigma Aldrich) and Trimethylalumium (TMA) (97%, Sigmal Aldrich) were used as metal precursors for $TiO_2$, $VO_x$, ZnO and $Al_2O_3$, correspondingly. To achieve a conformal coating on the MWCNTs forest, 20-30 seconds precursor pulses at a pressure of 0.4~0.5 mbar were applied. The pumping time was chosen as twice the pulse time to ensure sufficient evacuation of the residual precursor vapor and reaction products, thus avoiding chemical vapor deposition inside the films. After ALD coating, the samples were calcined with a ramp rate of 5$^o$C per min to remove the carbonaceous template. *In situ* x-ray diffraction (adapted Bruker D8 system) was used to characterize the crystallization behavior during the calcination. Scanning electron microscopy (SEM) accompanied with energy dispersive X-ray spectroscopy (EDXS) was performed using a FEI Quanta 200 F to resolve the morphology of the samples before and after annealing. TEM specimens were prepared by scraping off the films from the silicon substrate surface and suspending the resulting product in ethanol. A drop of this suspension was deposited on a carbon coated TEM grid. Bright-field TEM (BFTEM) and high-resolution TEM (HRTEM) images as well as selected area electron diffraction (SAED) patterns were collected using a FEI Tecnai F20 operated at 200 kV. Energy filtered TEM (EFTEM) elemental maps were obtained using a Philips CM30-FEG microscope operated at 300 kV. In addition, the crystallization temperature of metal oxide and oxidation temperature of MWCNTs were defined as $T_{cry}$ and $T_{ox}$, respectively.

## Results and discussion

Forests of MWCNTs with a thickness of ~6.5μm were coated with 100 cycles of $Al_2O_3$, $TiO_2$, ZnO and $VO_x$, respectively. After ALD, these samples were respectively calcined at 600 °C, 500 °C, 450 °C and 450°C in air for 3 hours to remove the carbon. In Fig. 1, SEM images show that all samples still exhibit a forest-like structure at the micro-scale upon calcination. The thicknesses of $Al_2O_3$, $TiO_2$ and $VO_x$ coated CNTs forests films are similar to the original thickness of the as-grown MWCNTs (~6.5 μm), but there is a noticeable shrinkage of the ZnO film (~4.7μm). With the exception of the minor changes in their thicknesses, there is no remarkable collapse observed in these films. The EDS results of the calcined samples shown in Fig. 1 (f)-(i) indicate that these metal oxide forests are free standing nanostructures without the carbon support. In the following, we will analyze their nano-scale morphology before and after calcination, and try to unveil the kinetics case by case using TEM and EDXS.

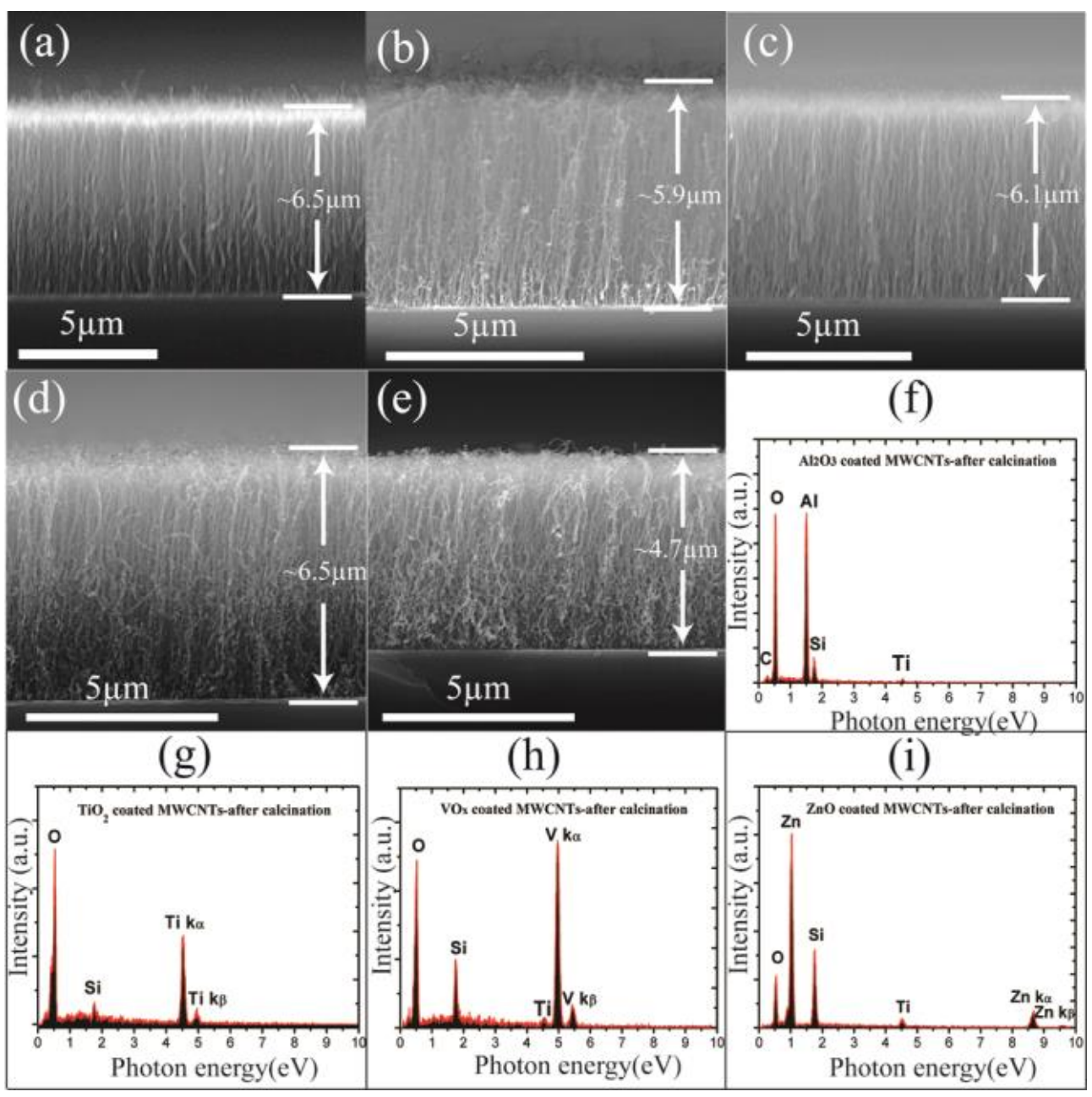

Fig.1 (a) shows the SEM image of $TiO_2$ coated MWCNTs before calcination. (b), (c), (d) and (e) SEM images show the $Al_2O_3$, $TiO_2$, $VO_x$ and ZnO coated MWCNTs after annealing in air. (f), (g), (h) and (i) show the EDXS results from (b), (c), (d) and (e), respectively.

### a. $Al_2O_3$ hollow nanotubes

Detailed TEM study combined with EFTEM elemental mapping applied to the as deposited $Al_2O_3$ coated MWCNTs sample shows that a uniform layer of amorphous $Al_2O_3$ was deposited on the surface of the MWCNTs (see Fig.2 (a)-(d)). The HRTEM image in Fig. 2(a) shows that the thickness of the $Al_2O_3$ coating on MWCNTs is smaller than a few nanometers. The uncoated section of the MWCNT indicated in Fig. (e) by a circle in the color-coded EFTEM elemental map might be due to overlapping of neighboring MWCNTs during the deposition which encumbered the exposure of this section to the precursors. After the deposition, the sample was annealed at 600°C in air for 3 hours to remove the MWCNTs. As can be seen in Fig. 2 (e), hollow $Al_2O_3$ nanotubes were formed successfully. The calcination process appears to have no effect on the tubular morphology, which is in agreement with an earlier work.[11]

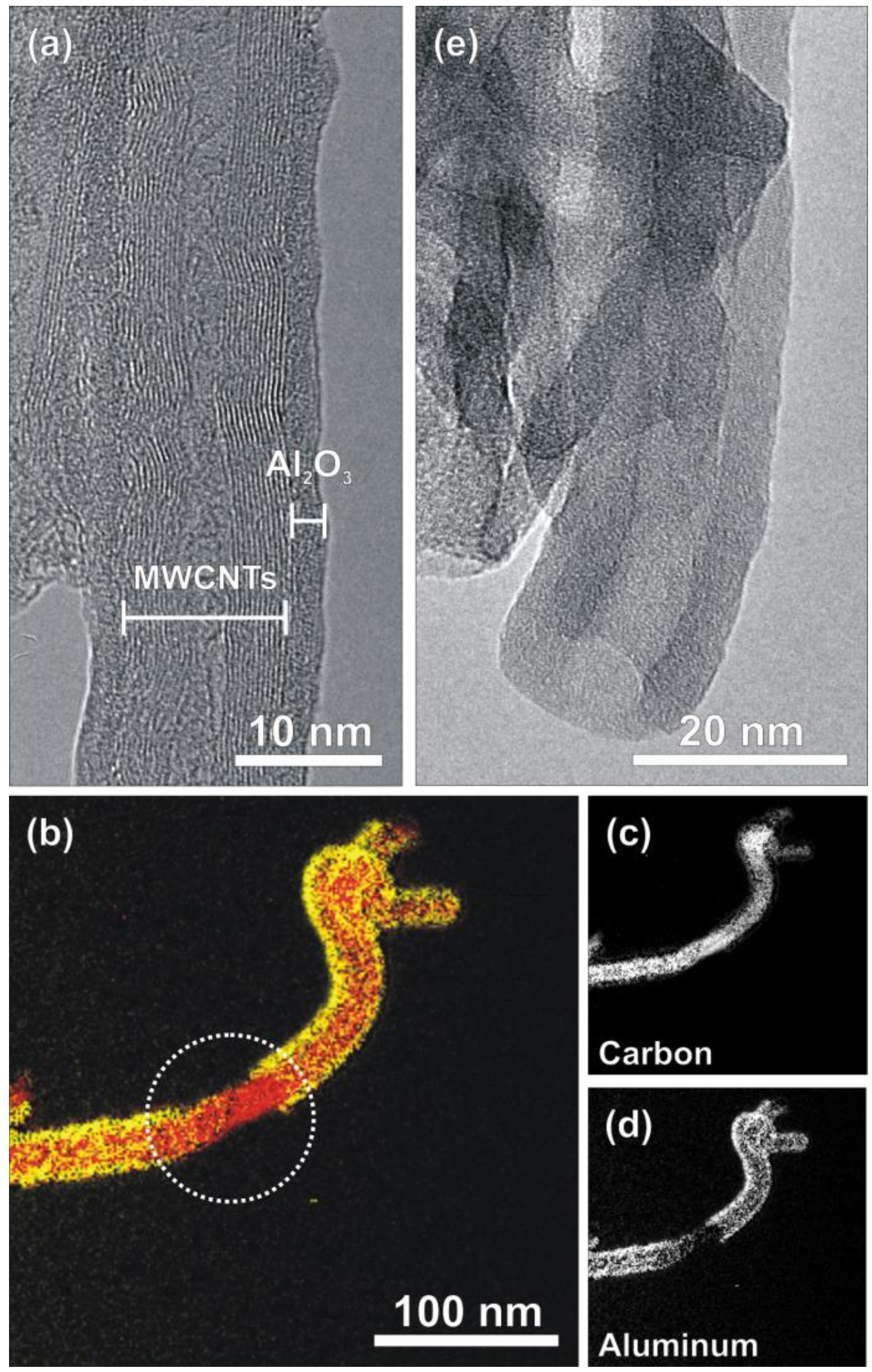

Fig.2 The amorphous $Al_2O_3$ coating and the graphitic layers of the MWCNTs are visible on the HRTEM image in (a). The carbon and aluminum contents of the nanotubes are presented in (b) showing the color-coded EFTEM map generated from individual (c) carbon (red) and (d) aluminum (yellow) elemental maps. A white circle in Fig. 2 (b) indicates a part that is not coated by $Al_2O_3$. TEM image of $Al_2O_3$ nanotubes after annealing in air at 600°C for 3 hours is given at (e).

### b. $TiO_2$ nanoparticle chains

Calcination of MWCNTs coated with continuous $TiO_2$ layers resulted in a remarkably different morphology at the nano-scale compared to MWCNTs coated with $Al_2O_3$ after the carbon removal. Fig. 3 (a) indicates that prior to calcination, a continuous layer of amorphous $TiO_2$ is present at the MWCNTs. After the calcination, the $TiO_2$ coating surrounding the MWCNTs transform into $TiO_2$ nanoparticles that are interconnected to each other. Fig.3 (b) indicates that these $TiO_2$ nanoparticles are fully solid and crystallized (e.g. not hollow). To understand the evolution of this transformation, detailed HRTEM characterization accompanied with chemical elemental mapping was carried out on samples that were quenched at different temperatures during calcination. According to the results of *in situ* XRD, the temperature of crystallization ($T_{cry}$) of ALD $TiO_2$ coating is ~400°C.[11] Thus, a series of samples were by taking quenches at 400, 425, 450 and 475°C for HRTEM characterization. In this manner, the morphologic transformation during calcination can be investigated. HRTEM images show that until 475 °C, the MWCNTs are still intact. However, for the sample quenched at 475°C (Fig. 3 (c)-(g)), the carbon content starts to decreases remarkably. As can be seen in Fig.3 (f) and (e), only a small amount of carbon is present inside the nanotube (indicated by white arrows). It should be here noted that the intense carbon signal observed at the carbon elemental map (indicated by red arrows) stems from the holey carbon of the TEM grid. On the other hand, the $TiO_2$ coating retains its tubular shape. This indicates that the carbon removal precedes the morphologic transformation. Fig. 3 (c) shows that even though the $TiO_2$ coating is still surrounding the partly removed MWCNTs, the coating starts to transform from amorphous to continuous and waved crystallized sections, with a rougher surface than the as deposited sample. As shown in Fig. 3 (d), some hollow $TiO_2$ nanotubes already appear.

XRD results on these samples further reveal that the $TiO_2$ coating started to crystallize at a temperature between 400 and 425°C and this crystallization proceeded more intensely at higher temperatures (Fig.4). It was expected that the calcination at 500°C for 3 hours is necessary to fully crystallize the $TiO_2$. From the HRTEM analysis, we know that the transformation of $TiO_2$ from tubular coating to nanoparticles started from ~475°C, which is higher than the oxidation temperature ($T_{ox}$) of MWCNTs, which will be discussed further later on. Upon exceeding this temperature, the crystallized $TiO_2$ sections began to sinter into interconnected nanoparticles when the MWCNTs were fully removed, causing the tubular morphology to collapse into chains of crystallized nanoparticles. Comparing with previous results of $Al_2O_3$ coated nanotubes, we can conclude that the lower $T_{cry}$ of $TiO_2$ allows the sintering of crystallized $TiO_2$ into nanoparticles instead of maintaining a tubular morphology. The thickness of the coating and the diameter of the MWCNT are also important parameters in determining the final morphology. When the coating is too thick, the merging of $TiO_2$ crystalline grains into nanoparticles is not likely to take place since the tubular coating is thick enough to maintain its own structure. Also, when the tube is too thick, it will be more difficult for the metal oxide walls to merge together.[7, 14, 15]

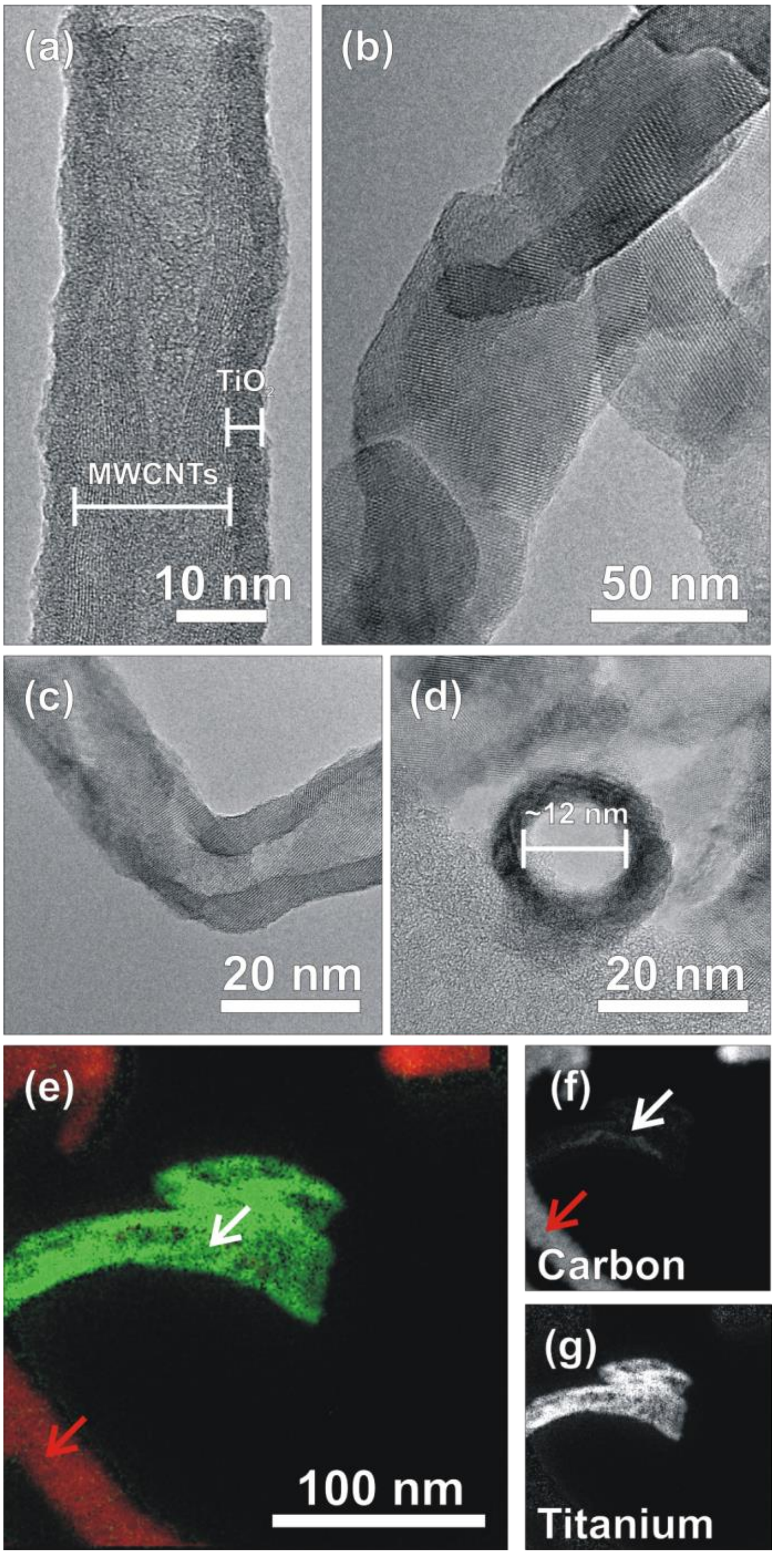

Fig.3 The amorphous $TiO_2$ coating and the graphitic layers of the MWCNTs are visible on the HRTEM image in (a). HRTEM images (b) of $TiO_2$ coated MWCNTs after calcination reveals the crystalline $TiO_2$ nanoparticle chains. (d) HRTEM images (c) and (d) of 475°C quenched sample show tubular morphology. The carbon (red) and titanium (green) contents of the nanotubes are presented in (e) showing the color-coded EFTEM map generated from individual (f) carbon and (g) titanium elemental maps.

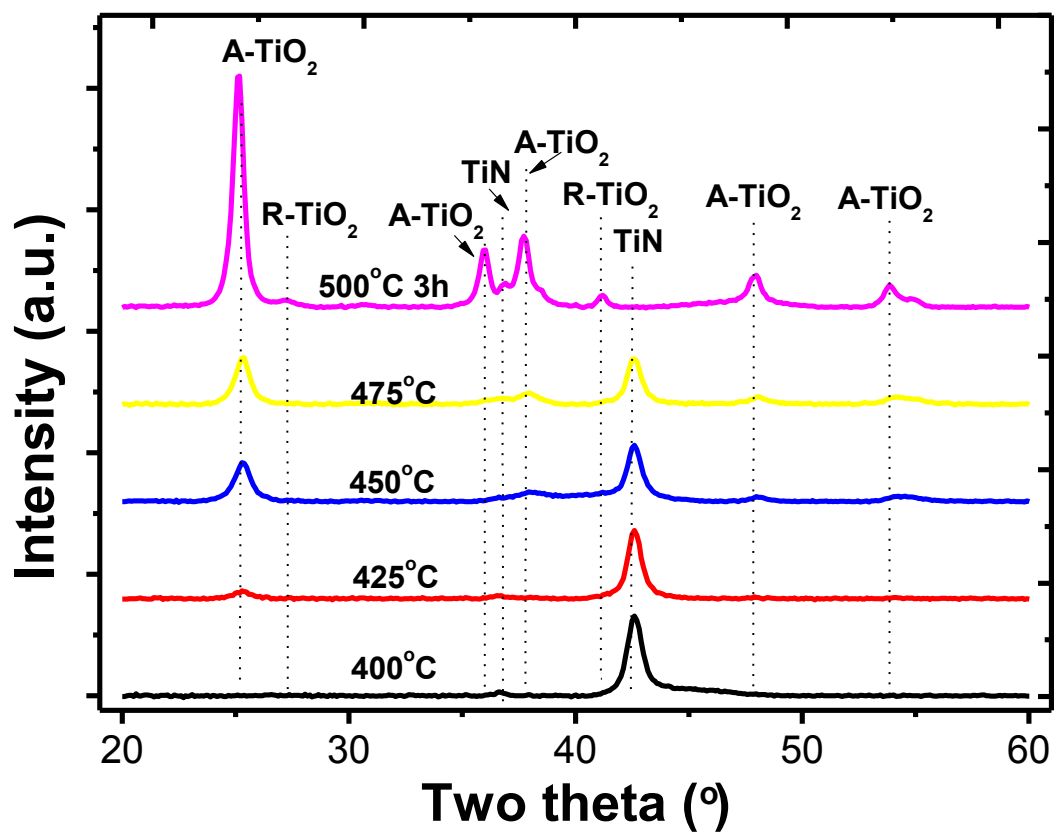

Fig. 4 XRD patterns of 400, 425, 450, 475°C quenched and 500°C calcined for 3 hours $TiO_2$ coated MWCNTs samples. A-$TiO_2$ and R-$TiO_2$ stand for $TiO_2$ of the anatase and rutile phase, respectively.

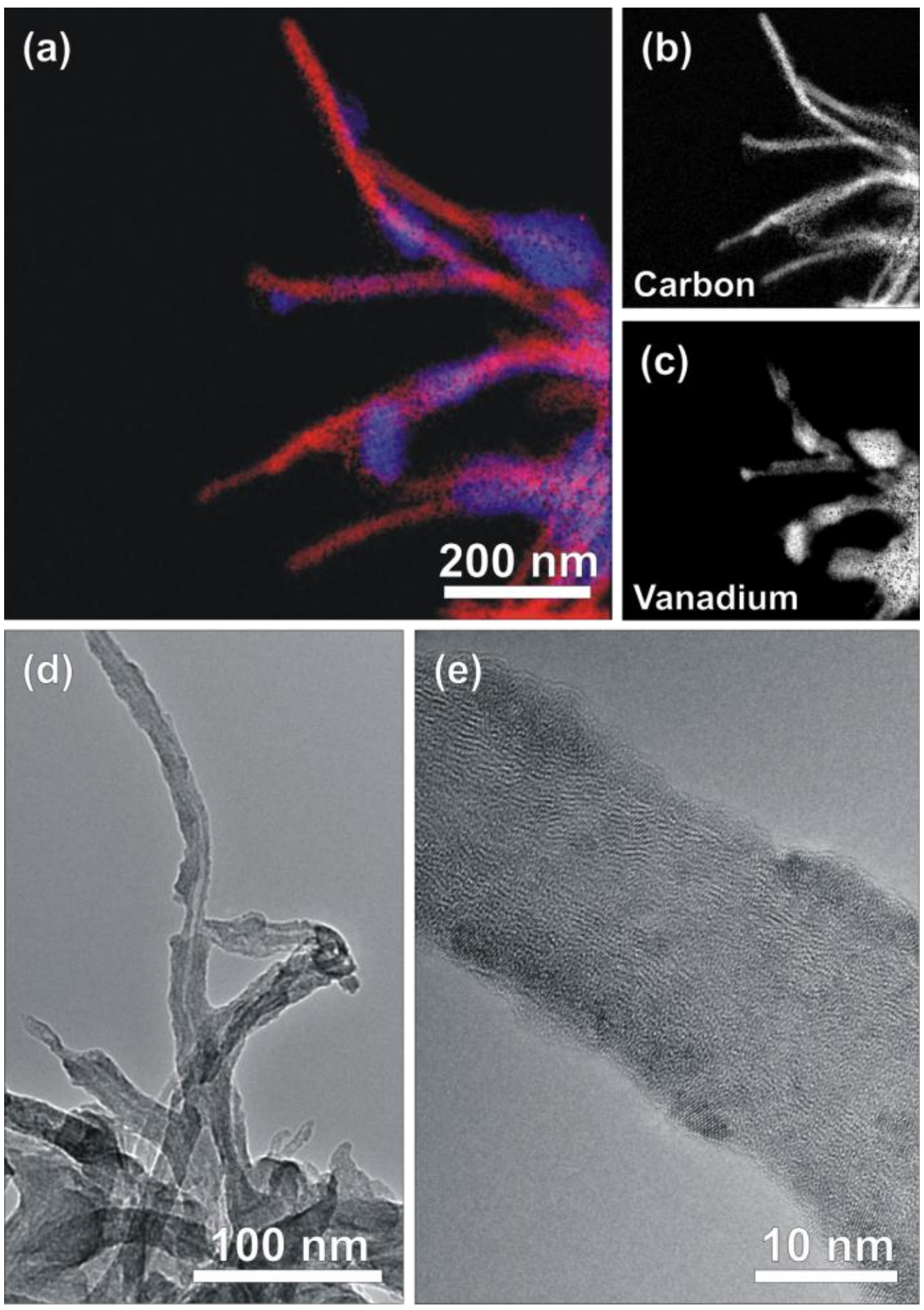

Fig. 5 The carbon and vanadium contents of the nanotubes are presented in (a) showing the color-coded EFTEM map generated from individual (b) carbon (red) and (c) vanadium (dark blue) elemental maps. TEM image at (d) depicts the uncoated and coated MWCNTs after ALD $VO_x$. (e) HRTEM image reveal crystalline $VO_x$ coatings in some spots prior to calcination (c).

**c. $V_2O_5$ and ZnO network of interconnected nanoparticles**

When applying ALD on MWCNTs to synthesize porous metal oxides, the homogeneity of the initial coating is usually considered to be crucial for preserving the whole structure after carbon removal.[16] If the metal oxide coating is discontinuous, the whole structure is generally expected to collapse after the carbon removal. However, the following results of synthesizing porous $V_2O_5$ and ZnO nanostructures using the same approach employed above show that a continuous and homogeneous coating is in fact not indispensable for structural preservation. HRTEM characterizations together with element mapping on 100 ALD cycles of $VO_x$ or ZnO coated MWCNTs show that ALD resulted in a non-uniform coating of $VO_x$ or ZnO on the surface of the MWCNTs, which is due to the inhomogeneity of the MWCNTs.[17, 18] Fig. 5 (a)-(c) show $VO_x$ layers on parts of the MWCNTs, whereas other sections remained uncoated (Fig. 5 (b)). The HRTEM image in Fig. 5 (e) shows that in some regions, the as deposited $VO_x$ coating has already crystallized into the $V_2O_5$ phase. ALD of ZnO on MWCNTs shows similar results. ZnO nanoparticles are attached to the surface of MWCNTs (see Fig. 6 (a)-(d)). The coverage of ZnO on the MWCNTs is even less than for $VO_x$. Fig. 6 (d) shows that the ZnO nanoparticles are already crystallized as deposited.

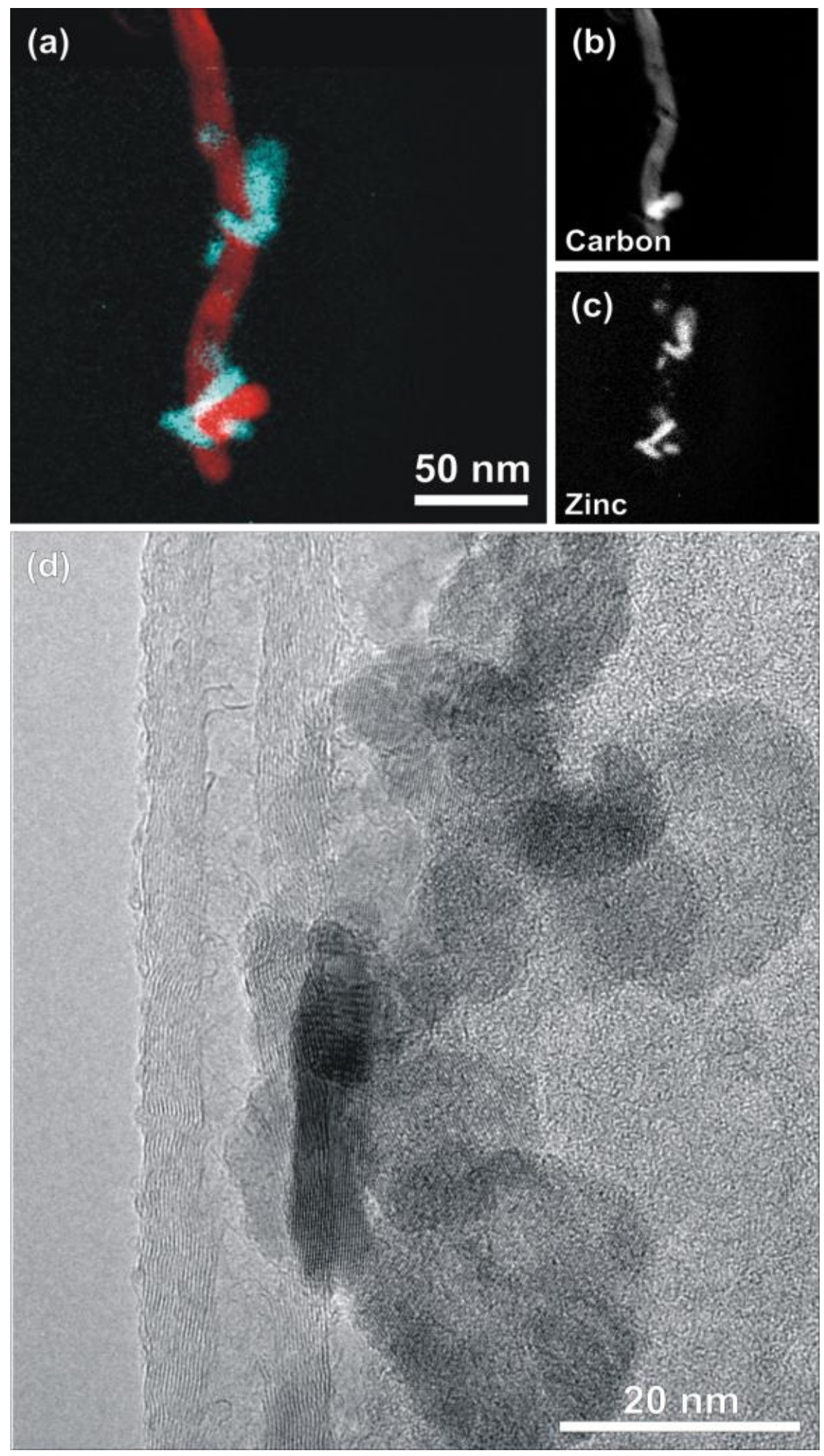

Fig. 6 The carbon and zinc contents of the nanotubes are presented in (a) showing the color-coded EFTEM map generated from individual (b) carbon (red) and (c) zinc (light blue) elemental maps. (d) HRTEM image reveals crystallized ZnO nanoparticles attached to the surface of MWCNTs.

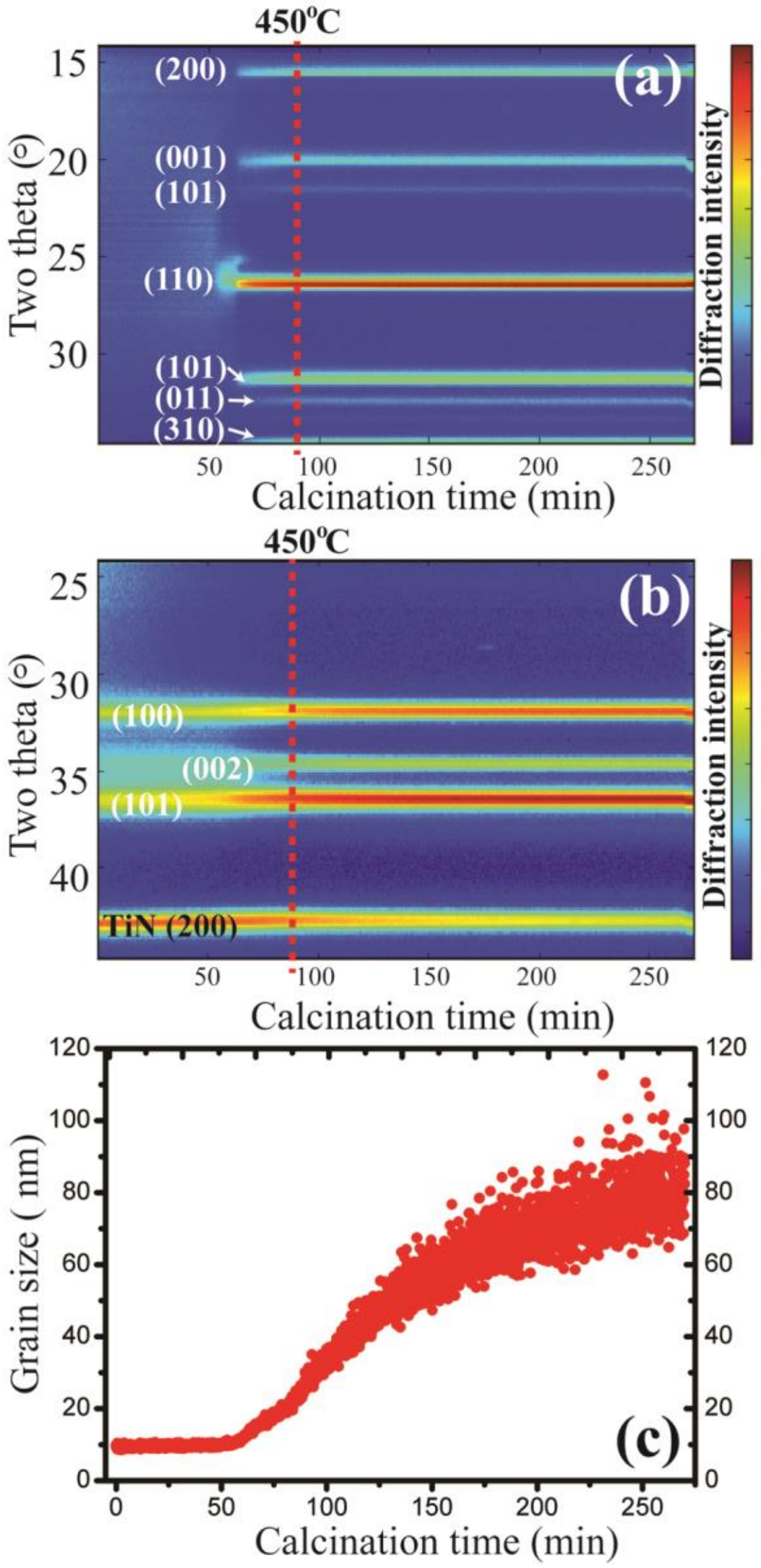

Fig.7 *In situ* XRD of $VO_x$ (a) and ZnO (b) coated MWCNTs annealed from 20 to 450°C with a ramp rate of 5°C per min and stay at 450°C for 3 hours. The grain size of ZnO versus calcination time is shown in (c), calculated from in situ XRD data using the Scherrer equation.

In order to remove the carbon, these samples were calcined at 450°C in air for 3 hours. SEM images in Fig. 1 (e) and (f) show that after the calcination, the $VO_x$ and ZnO samples still exhibit a similar forest like morphology at the micro-scale. The $VO_x$ forest has the same overall thickness of ~6.5μm as the original MWCNTs, whereas the ZnO slightly shrinks to ~4.7μm. However, in spite of the particle-type nature of the metal oxide ALD coating, the structures do not collapse or disintegrate upon removal of carbon.

*In situ* XRD was used to monitor the crystallization behavior of these metal oxides when calcining these films from room temperature to 450°C. $VO_x$ crystallized into $V_2O_5$ at approximately $T_{cry}$=290°C and showed no increase in grain size after crystallization (Fig. 7 (a)), whereas ZnO coatings did not show remarkable changes of the diffraction peaks during calcination since they were already crystallized as deposited (Fig.7 (b)). However, a notable intensification of the diffraction peaks of ZnO (100), (002) and (101) was observed in Fig.7 (b), which implies that there was an increase in the grain size of the ZnO coating. Basing on the evolution of the diffraction peak at (101), we used the Scherrer equation to plot the variation of grain size versus calcination time, as shown in Fig.7 (c). The result shows that the initial grain size of the as deposited ZnO was about 10nm and started to remain stable when the temperature reaches ~270°C. As the calcination temperature increases, the ZnO grains grow, which implies a merging process of ZnO nanoparticles during the calcination.

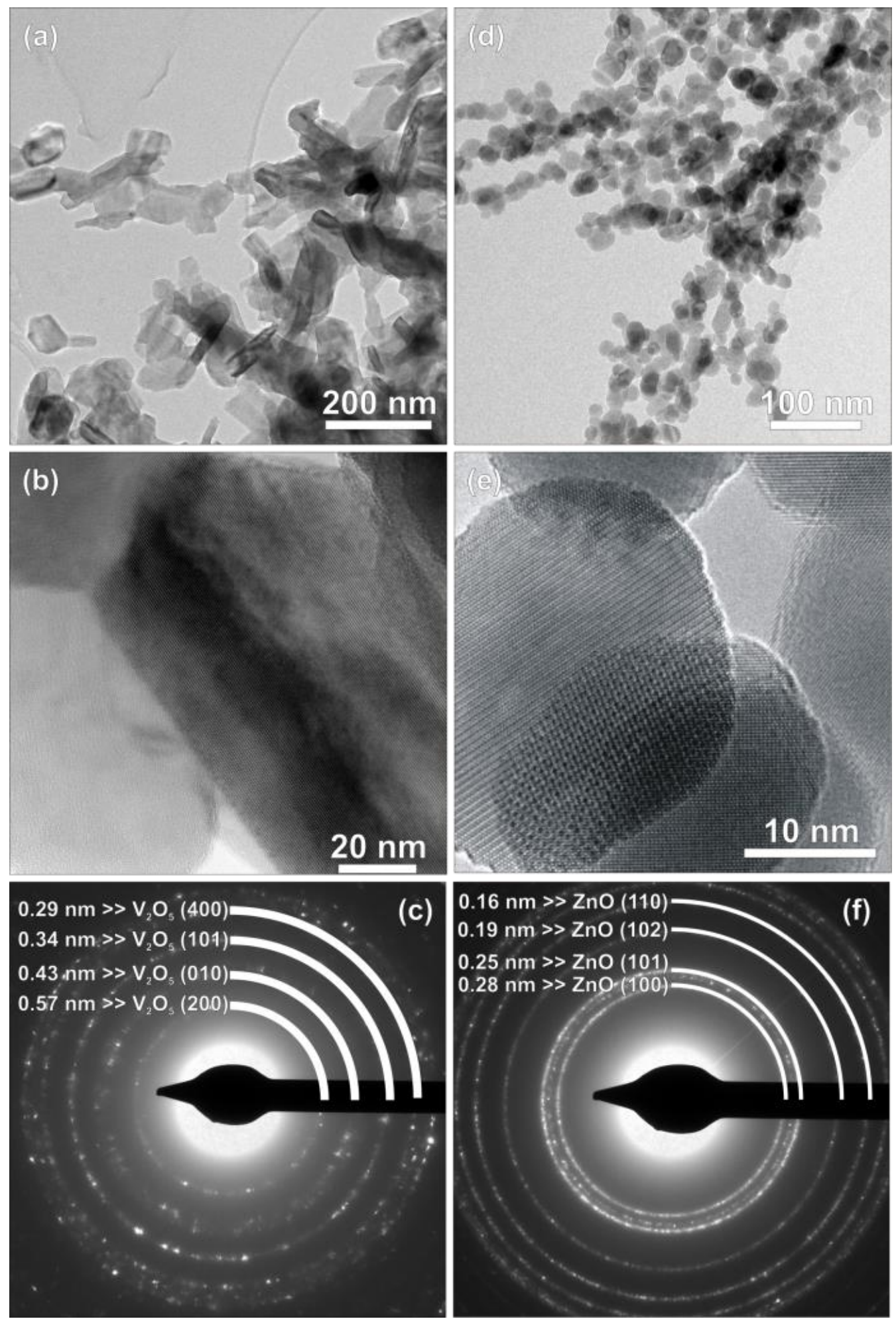

Fig.8 HRTEM images of the interconnected $V_2O_5$ nanorods after annealing $VO_x$ coated MWCNTs in air (a-b). SAED pattern of crystallized $V_2O_5$ is shown in (c). HRTEM images of the interconnected ZnO nanoparticles after annealing ZnO coated MWCNTs in air (d) and (e). SAED pattern of crystallized ZnO is shown in (f).

To investigate the morphology of the $VO_x$ and ZnO coatings after the calcination, HRTEM was carried out for these two films. Fig. 8 (a)-(b) show that the $VO_x$ coating crystallizes into interconnected $V_2O_5$ nanrods, with diameter of 20 to 80nm and lengths of 60nm to more than 100nm. The electron diffraction rings in the selected area electron diffraction (SAED) pattern of crystallized $VO_x$ shown in Fig. 8 (c) is indexed according to the various planes of the $V_2O_5$ phase. In the case of ZnO, as shown in Fig. 9 (a) and (b), spherical ZnO nanoparticles with diameters from 10-50nm connect to each other in a similar manner as the $V_2O_5$ nanorods, indicating that during the carbon removal the initial separate nanoparticles connected to each other and grew into larger particles. The SAED pattern in Fig. 9 (c) confirms that these ZnO nanoparticles are crystalline. The coverage of ZnO nanparticles on MWCNTs is lower in comparison to the coverage for $VO_x$, explaining the shrinkage in the thickness (from ~6 to ~4.7μm) of the film after calcination compared to the film of $V_2O_5$ nanorods, as shown in Fig.1. Even though a uniform coverage of metal oxide on the MWCNTs may facilitate structure preservation during carbon removal, the examples of $VO_x$ and ZnO clearly demonstrate that calcination of samples with a non-continuous coating can also result in a supportless nanostructure of porous metal oxide through merging and sintering of the separate nanoparticles into 3D interconnected networks during the calcination.

### d. Kinetics of carbon removal

We systematically investigated the oxidation temperatures of MWCNTs coated with $Al_2O_3$, $TiO_2$, ZnO and $VO_x$ by using EDXS to record the carbon signal at different quenches of the samples during calcination, as shown in Fig. 9 and 10. In Fig. 9, MWCNTs were coated with 100 ALD cycles of $Al_2O_3$, $TiO_2$, ZnO and $VO_x$ respectively and annealed from room temperature to 600°C. Several quenches at different temperatures with 50°C intervals were preferred for EDXS measurement. For pure MWCNTs without metal oxide coating, $T_{ox}$ is found to be ~500°C. Upon coating with metal oxides, $T_{ox}$ of MWCNTs changes accordingly. $TiO_2$ and $VO_x$ coated MWCNTs have a ~50°C decrease in $T_{ox}$ whereas for ZnO coated MWCNTs $T_{ox}$ shows a ~100°C decrease.[16, 19] However, a ~50°C increase of $T_{ox}$ is observed for the $Al_2O_3$ coated samples.

Different numbers of ALD cycles were also applied to investigate the effect of the thickness of the metal oxide coating on $T_{ox}$. In Fig. 10, the carbon signal measured by EDS for 50, 100 and 200 cycles $TiO_2$ and $Al_2O_3$ coated MWCNTs are shown. For $TiO_2$ coated MWCNTs, the $T_{ox}$ are independent of the thickness of $TiO_2$. For the $Al_2O_3$ coated MWCNTs, the thicker the $Al_2O_3$ layer, the higher the $T_{ox}$. For 200 cycles of $Al_2O_3$, removal of the carbon is not yet complete, even though the quench is taken at 650°C. The decrease in $T_{ox}$ of MWCNTs can be attributed to the catalytic effect of coated metal oxides. Aksel et al proposed a possible mechanism of explaining this catalytic effect.[16] First the oxygen in the metal oxide reacts with the carbon of the MWCNTs and generates CO and an oxygen vacancy in the metal oxide layer. Then oxygen vacancies will migrate to the gas-solid interface and be occupied with oxygen from the ambient. The results shown in Fig.9 are in agreement with this hypothesis. However, the results in Fig.10 show that the uniformity of the thin film coating also influences the oxidation temperature. If the layer of metal oxide is not dense enough e.g. $TiO_2$ compared to $Al_2O_3$, increasing the thickness does not increase the oxidation temperature of the MWCNTs, since pin holes probably allow oxygen vacancies to migrate to the gas-solid interfaces. Comparing Fig.2 (a) with Fig. 3(a), the $TiO_2$ coating on the MWCNTs is less uniform than the $Al_2O_3$, which implies that the uniformity and the quality of $Al_2O_3$ coating is better than $TiO_2$ and would have fewer pin holes. On the other hand, compared to other methods of coating metal oxides on MWCNTs, ALD is a well-known technique for growth of higher quality of metal oxides. Given perfect starting surface, ALD will grow metal oxide layer with limited pinholes. An increase of oxidation temperature was found when $TiO_2$ was deposited by ALD on oxygen plasma treated MWCNTs, which have more defective surface and are easier for $TiO_2$ to nucleate and form ALD coating with higher quality.[18]

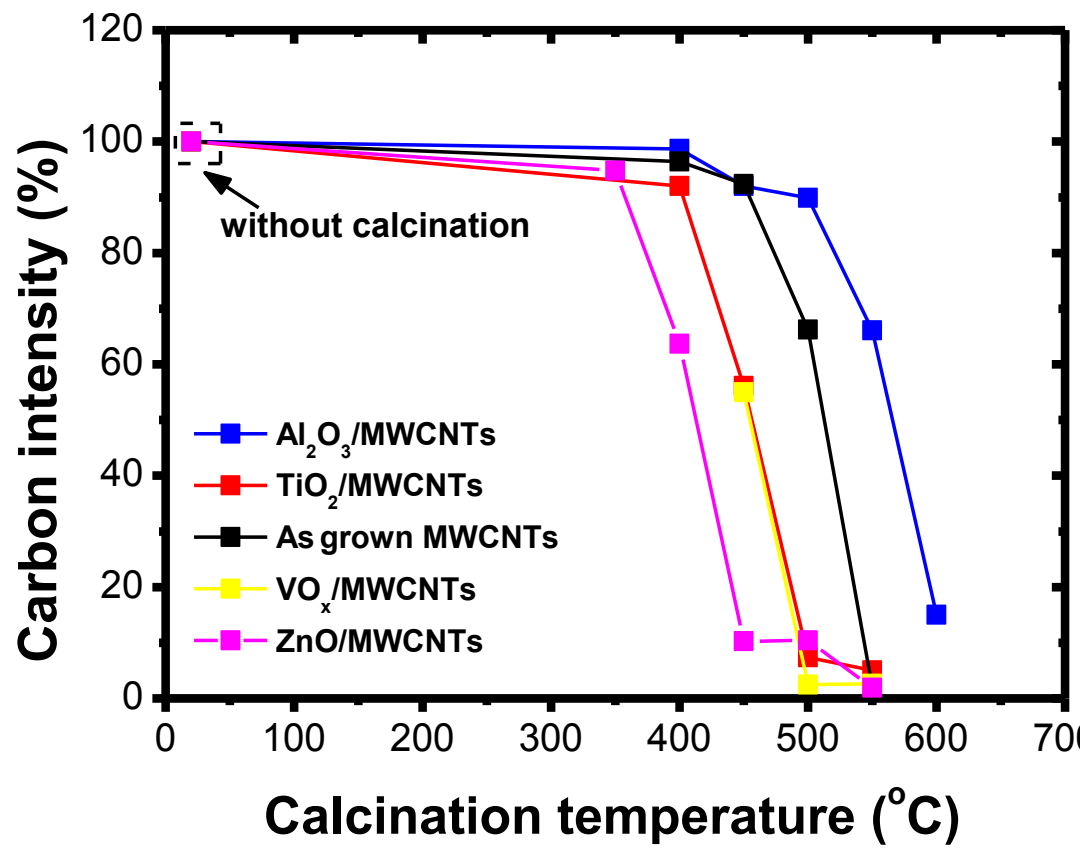

Fig. 9 EDXS measured carbon contents of $Al_2O_3$, $TiO_2$, $VO_x$ and ZnO coated MWCNTs with 100 ALD cycles quenched at different temperatures. The result of referenced uncoated MWCNTs is also shown.

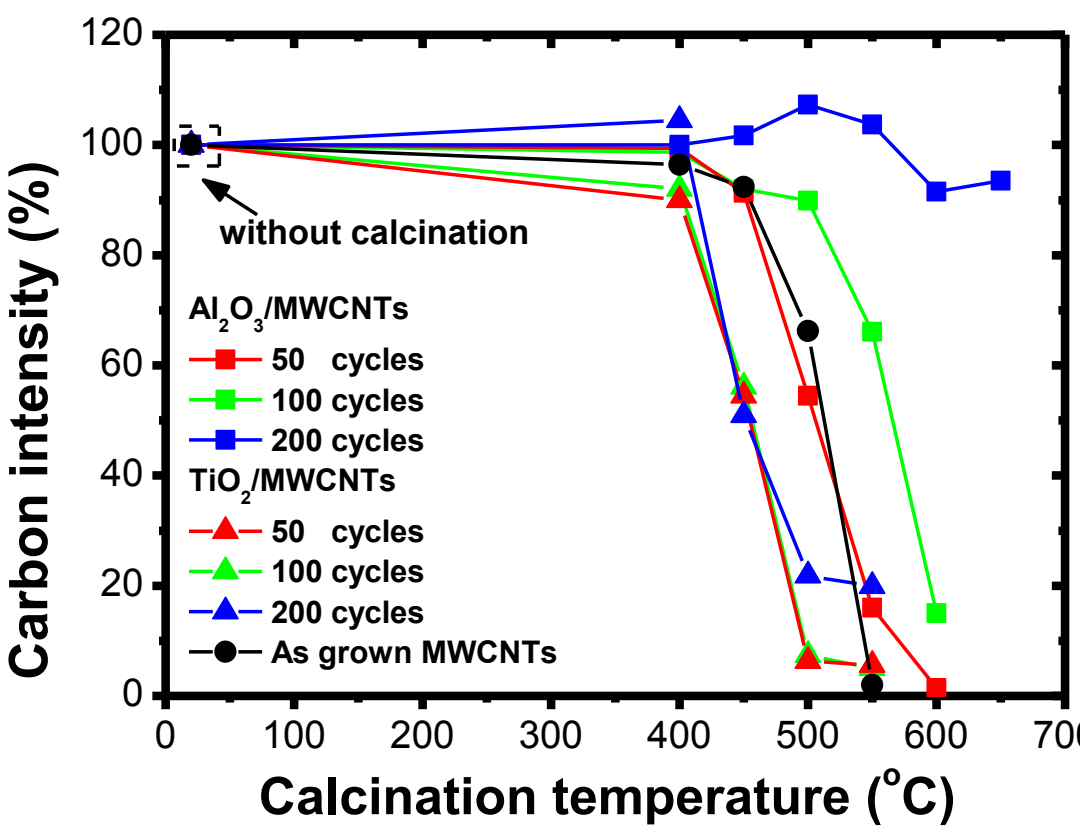

Fig.10 EDXS measured carbon contents of $Al_2O_3$ and $TiO_2$ coated MWCNTs with 50, 100 and 200 ALD cycles quenched at different temperatures. The result of referenced uncoated MWCNTs is also shown.

## Conclusion

To summarize, by combining ALD metal oxides on MWCNTs and subsequent calcination, porous nanostructured metal oxides were successfully synthesized. A schematic illustration combined with TEM images is given in Fig. 11. The process of morphologic transformation during calcination is sketched according to $T_{cry}$ and $T_{ox}$, as indicated in the scheme. From the systematic investigation above we can conclude that:

i. For crystalline oxides (i.e. when calcining at temperatures higher than $T_{cry}$), sintering becomes a key mechanism during the calcination and determines the final morphology. For the $TiO_2$ sample, the driving force to reduce the surface area in combination with a sufficient mobility of the atoms causes a collapse of the hollow tube nanostructure into chains of anatase nanoparticles.

ii. As illustrated by the examples of $VO_x$ and ZnO, a continuous coating of the MWCNTs is not indispensible for avoiding the entire structure to collapse during the removal of the carbon support. Self-supported, interconnected 3D structure can be formed by merging and sintering of the initial separate nanoparticles during the calcination.

iii. The uniformity and thickness of the metal oxide coating affects the $T_{ox}$ of the MWCNTs. Thick layers of metal oxide with high quality increase the $T_{ox}$, which is due to the retardance of oxygen vacancy diffusion from carbon-metal oxide interface to the exterior surface.

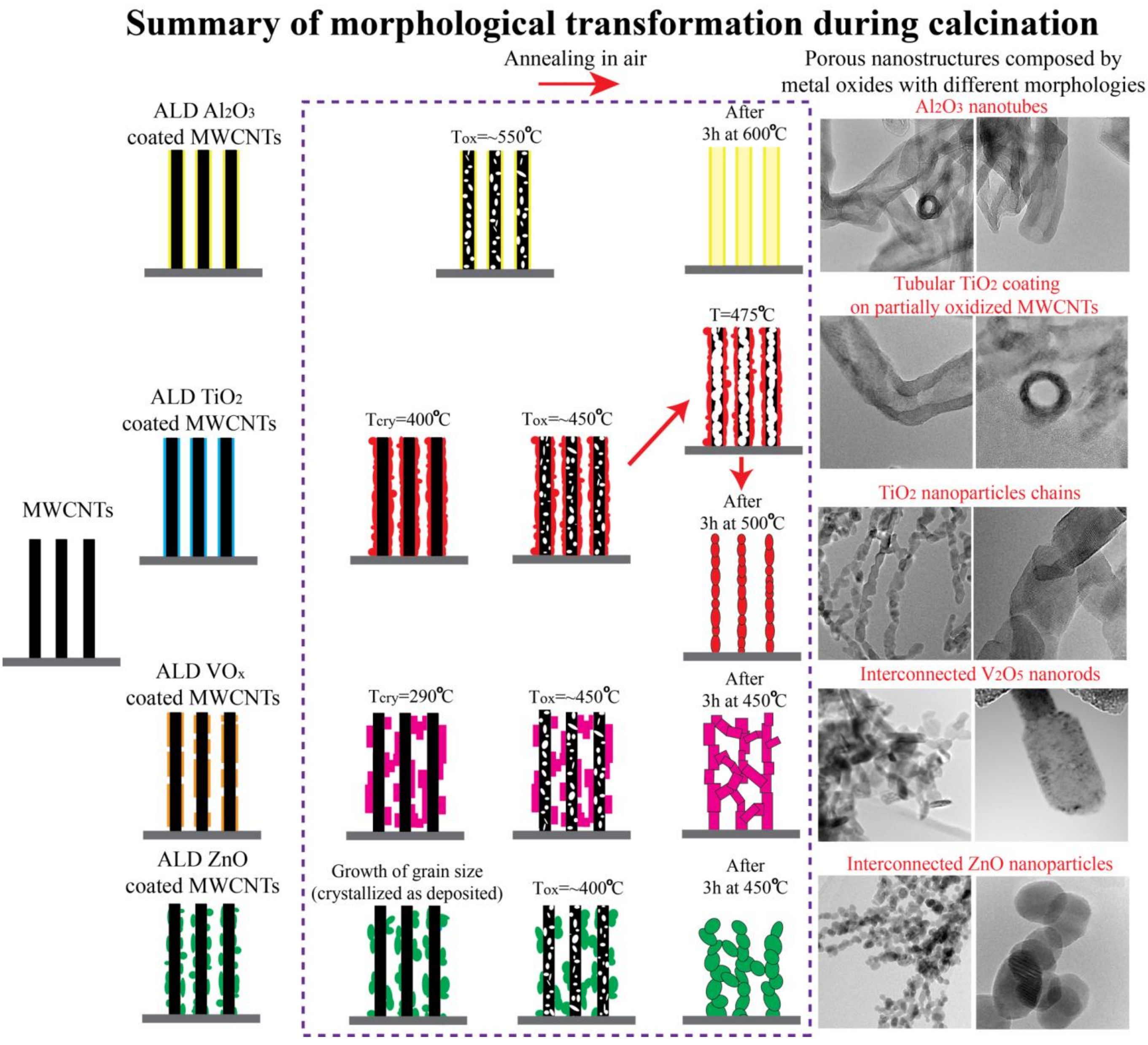

Fig. 11 Schematic demonstration of the formation of porous metal oxide nanostructures by ALD on MWCNTs followed by calcination.

## Acknowledgements

The authors wish to thank the Research Foundation - Flanders (FWO) for financial support. The authors acknowledge the European Research Council for funding under the European Union's Seventh Framework Programme (FP7/2007-2013)/ERC grant agreement N°239865-COCOON and N°335078-COLOURATOMS. The authors would also want to thank the support from UGENT-GOA-01G01513 and IWT-SBO FUNC (STM-Vlaanderen).

## Notes and references

[a]Department of Solid State Sciences, Ghent University, Krijgslaan 281/S1, B-9000 Ghent, Belgium

[b]Department of Physics, Electron Microscopy for Materials Science (EMAT), University of Antwerp, Groenenborgerlaan 171, B-2020 Antwerp, Belgium

[c]IMEC, Kapeldreef 75, B-3001 Leuven, Belgium

‡These authors contributed equally.